\documentstyle[12pt,a4,amssymb,times]{article}

\def\a{\alpha}
\def\b{\beta}
\def\th{\theta}
\def\S{\Sigma}
\def\s{\sigma}

\def\D{\Delta}
\def\e{\epsilon}
\def\L{\Lambda}

\def\c{\chi}
\def\g{\gamma}

\def\Om{\Omega}
\def\p{\pi}
\def\d{\delta}
\def\m{\mu}
\def\n{\nu}
\def\l{\lambda}

\def\pa{\partial}
\def\to{\rightarrow}
\def\cs{{\cal S}}
\newcommand{\be}{\begin{equation}}
\newcommand{\ee}{\end{equation}} 
\newcommand{\bea}{\begin{eqnarray}}
\newcommand{\eea}{\end{eqnarray}}

\begin{document}

\begin{titlepage}

\begin{flushright} 
IST/DM/16/00
\end{flushright}

\bigskip
\bigskip
\bigskip
\bigskip

\begin{center}

{\bf{\Large Super Chern-Simons Theory and Flat Super Connections on a Torus }}

\end{center}
\bigskip
\begin{center}
 A. Mikovi\'c \footnote{E-mail address: amikovic@math.ist.utl.pt. On leave of 
absence 
from Institute of Physics, P.O.Box 57, 11001 Belgrade, Yugoslavia}
and R. Picken \footnote{E-mail address: rpicken@math.ist.utl.pt}
\end{center}
\begin{center}
\it
Departamento de Matem\'atica and Centro de Matem\'atica Aplicada, 
Instituto Superior 
Tecnico, Av. Rovisco Pais,
1049-001 Lisboa, Portugal

\end{center}

\normalsize 

\bigskip 
\bigskip
\begin{center}
                        {\bf Abstract}
\end{center}    
We study the moduli space of a super Chern-Simons theory on a 
manifold with the topology ${\bf R}\times \S$, where $\S$ is 
a compact surface. The moduli space is that of flat super connections modulo 
gauge transformations on $\S$, and we 
study in detail the case when $\S$ is atorus and the supergroup is $OSp(m|2n)$.
The bosonic moduli space is determined by the flat connections for the maximal
bosonic subgroup $O(m)\times Sp(2n)$, 
while the fermionic moduli appear only for special parts of the
bosonic moduli space, which are determined by a vanishing determinant of a 
matrix associated to the bosonic part of the holonomy. If the CS
supergroup  is the exponential of a super Lie algebra, the fermionic moduli 
appear
only for the bosonic holonomies whose generators
have zero determinant in the fermion-fermion block of the super-adjoint
representation. A natural
symplectic structure on the moduli space is induced by the
super Chern-Simons theory and it is determined by the Poisson bracket algebra
of the holonomies. We show that the symplectic structure of 
homogenous connections is useful for understanding the properties of
the moduli space and the holonomy algebra,
and we illustrate this on the example of the $OSp(1|2)$ supergroup.

\end{titlepage}
\newpage

\section{Introduction}

The study of Chern-Simons (CS) theories based on non-compact groups is 
important 
for understanding classical and quantum properties of three-dimensional (3d) 
gravity theories \cite{car}. One can show that pure 3d general relativity is 
$ISO(1,2)$ (3d Poincare group) CS theory, while  3d general relativity
with a cosmological constant $\L$ is a CS theory based on
the group $SO(2,2)$ for $\L < 0$ (Anti-de-Sitter gravity), while for $\L >0$
it is the $SO(1,3)$ CS theory (de-Sitter gravity) \cite{w,at}. Furthermore, 
3d supergravity is a super 
CS theory for the 3d super-Poincare group \cite{w}, AdS supergravity is a 
super CS for a certain class of supergroups \cite{gst,hms}, one of which is 
$OSp(m|2)\times OSp(n|2)$ \cite{at}, 
while for de-Sitter supergravity the relevant group is $OSp(1|2;{\bf C})$
\cite{kmvw}. 

The moduli space of a CS theory represents the physical degrees of freedom 
(dof)
and it is important for understanding the classical solutions and for the 
quantization. When the three-manifold has topology ${\bf R}\times \S$, where
$\S$ is compact, then the moduli space for compact groups is well understood
\cite{ccs}. However, when the CS group is non-compact, there are few results.
The 3d Poincare group case has been studied in \cite{ml} when $\S$ is a torus
$T^2$,
and the moduli space is finite dimensional with a non-Hausdorf 
topology. The super-Poincare case for a torus has been studied in \cite{wmn}, 
and the novelty is that the fermionic moduli only appear for special parts
of the bosonic moduli space. Also the structure of the 3d super-Poincare 
algebra is such that there are no quadratic fermion contributions to the
bosonic curvature two-form, so that the problem of constructing the moduli 
space is simpler than in the case of other relevant super groups.

Introducing a symplectic structure on the moduli space is important for 
understanding solutions and quantization. In the case of groups relevant for 
3d gravity, one can use the
approach of \cite{th}, where traces of holonomies are used as
basic variables. This was extended to the super de-Sitter case in 
\cite{urr}. However, the problem with using traces of holonomies is that
it is difficult to understand the fermionic moduli space, basically because
the trace is an even element of the Grassman algebra while fermionic moduli
belong to the odd part. Hence it is better to consider the holonomy
matrix elements, and their algebra in order to find fermionic
observables. Even in the purely bosonic case there are advantages in
working with the holonomy matrix elements \cite{np}.

In order to explore theses issues we consider a super CS theory for 
the $OSp(m|2n)$ group on $R \times T^2$. In section 2 we give a definition
of a super CS theory for an arbitrary super Lie group and a three manifold
$M$. In section 3 we specialize to the case $M = R\times \S$ and
describe the canonical formalism. In section 4 we 
introduce the holonomies for a  super Lie group and describe how to determine
the moduli space for $\S = T^2$. In section 5 we study the moduli space
for the $OSp(m|2n)$ supergroup and we determine the condition for the 
existence of the fermionic moduli. In section 6 we discuss the basic features
of the symplectic structure on the moduli space, and show how the homogeneous
connections can be used to extract information about the moduli space in the
exponential sector of the moduli space. We extend these results
to the case when the CS supergroup is the exponential of an arbitrary super Lie
algebra. In section 7 we work out the $OSp(1|2)$ case in detail and we
present our conclusions in section 8.

\section{Super Chern-Simons Theory}

Let $g_s$ be a super Lie algebra, with a basis $ T_I =\{ J_a , Q_\a \}$,
where $J_a$ are even elements and $Q_\a$ are odd elements. $g_s$ 
has a super Lie bracket $[,\}$, which satisfies
\be
[X,Y\} = (-1)^{|X||Y|+1}[Y,X\}\quad, \label{slb}
\ee
as well as the super Jacobi identity 
\be
[X,[Y,Z\}\} =  [[X,Y\},Z\} +(-1)^{|X||Y|} [Y, [X,Z\}\} \quad, 
\ee
where $X$, $Y$ and $Z$  are elements of definite parity 
($|X|= 0,1$ for an even, odd element respectively). The super Lie algebra
$g_s$ is determined by the super Lie brackets of the basis elements
\be
[ T_I,T_J \} = {f_{IJ}}^K T_K ,\label{def} 
\ee
where ${f_{IJ}}^K$ are the structure constants, belonging to $\bf R$ or 
$\bf C$.

The algebra $g_s$  can be represented by 
matrices so that the relations (\ref{def}) take
the following form
\bea
[ J_a ,J_b ] &=& {f_{ab}}^c J_c \quad,\\
{[} J_a ,Q_\a ] &=& {f_{a \a}}^\b Q_\b  \quad,\\ 
 \{ Q_\a ,Q_\b \} &=& {f_{\a\b}}^a J_a \quad,\label{esr}
\eea
where $[X,Y] = XY-YX$ and $\{ X,Y \} = XY +YX$.

Let $B_N$ be a Grassmann algebra generated by $\{\th_1, ...,\th_N \}$, which
satisfy
\be
\{ \th_i ,\th_j \} =0 \quad 1\le i,j \le N \quad.\label{gar}
\ee
The natural basis of $B_N$ can be split into an even and odd part
\bea
E_N &=& \{ 1, \th_i \th_j , \th_i \th_j \th_k \th_l , ... \} \\
O_N &=& \{ \th_i, \th_i \th_j \th_k , ... \} \quad,
\eea
so that $B_N = L(E_N) \oplus L(O_N)=B_N^+ \oplus B_N^-$. We can now define a 
Grassmann enveloping algebra of $g_s$, $B(g_s)$, as
\be
B(g_s)= \{ X = X^a J_a + \chi^\a Q_\a \, |\, X^a \in B_N^+ \, ,\,
\chi^\a \in B_N^- \,\}
 \quad.\label{gvr}
\ee

Consider now a one-form $A$ on a three-manifold $M$, taking values in $B(g_s)$.
If $\{ x^\m \}$ are coordinates on $M$, then
\be
A = A_\m^a (x)dx^\mu J_a + \psi_\m^\a (x)dx^\mu Q_\a   = A_\m^I (x)dx^\mu  T_I
\quad.
\ee
Note that in physics there is a restriction $A_\m^a \in B_0 = \bf R$ or 
$\bf C$ while $\psi_\m^\a \in B_1 = L(\th_i)$.
The super Chern-Simons theory associated to a super Lie group $G_s$, whose
Lie algebra is $B(g_s)$, is
defined by the action functional
\be
S[A] = \int_M str \left( AdA + {2\over 3}A\wedge A \wedge A \right) \quad,
\label{scs}
\ee
where $d$ is the exterior derivative, and $str$ is the super trace
(invariant bilinear form on $g_s$), given by
\be
str(J_a J_b) = \eta_{ab}= \eta_{ba} \quad,\quad str(Q_\a Q_\b) = C_{\a\b}= 
-C_{\b\a}\quad str(J_a Q_\a ) = 0 \quad.\label{str}
\ee 
This is denoted as
\be
str(T_I T_J) = \eta_{IJ} \quad.
\ee

The equations of motion are the extremal points of $S[A]$
\be
{\d S\over \d A} = dA + A\wedge A = F = 0 \quad,\label{fcm}
\ee
which are satisfied by the flat $G_s$ connections on $M$. This can be written 
in a more explicit form
\bea
F^a &=& dA^a + {f_{bc}}^a A^b \wedge A^c + {f_{\a\b}}^a \psi^\a \wedge \psi^\b
 =0 \quad,
\nonumber\\
F^\a &=& d\psi^\a + {f_{a\b}}^\a A^a  \wedge \psi^\b = 0 \quad.
\eea 

The super CS theory is invariant under the gauge transformations 
$g: M \to G_s$, so that the connection $A$ and the connection
\be
\tilde A = g A g^{-1} + dg g^{-1} \label{gt}
\ee
are equivalent. Hence the moduli space is the space of flat connections 
modulo gauge transformations.

\section{$M={\bf R}\times \S$ case}

We study now the case when $M = {\bf R}\times \S$. In this case there is a
natural split of the coordinates $x^\m = \{t,x^i \}$ where $t\in \bf R$ and
$x^i$ are coordinates on $\S$. This induces the split for the one forms
$A_\m = \{ A_0 , A_i \}$, so that the action can be written in the Hamiltonian
form
\be
S = \int_{t_1}^{t_2}dt\int_\S d^2 x \left[ E_I^i {\dot A}^I_i - A_0^I G_I
- \l^I_i \th^i_I \right] \quad,\label{hf}
\ee
where $G_I$ are the first-class Gauss constraints
\be
G_I = \pa_i E^i_I + {f_{IK}}^J A_i^K E_J^i = 0 \quad,\label{gc}
\ee
while $\th^i_I$ are the second-clsss constraints
\be
\th^i_I = E_I^i - \eta_{IJ}\e^{ij}A_j^J =0 \quad.\label{sc}
\ee
Here $E_I^i$ is the canonically conjugate momentum to $A_i^I$ and $\e_{ij}$ 
is the antisymmetric tensor density
on $\S$. One can introduce the super Poisson brackets (PB) for the functions 
on the phase space $(E,A)$ \cite{bf}
\be
\{F(x),G (y)\}_{PB} = \int_\S d^2 z \left({\d F (x)\over\d E^i_I (z)}
{\d G (y)\over\d A_i^I (z)} -
(-1)^{|A_i^I||E_I^i|}{\d F (x)\over\d A_i^I (z)}{\d G (y)\over\d E^i_I (z)}
\right)\quad, \label{pob}
\ee 
so that
\be
\{ E_I^i (x) , A_j^J (y) \}_{PB} = \d^i_j \d_I^J \d (x-y) \quad.
\ee 

The first-class constraints form a closed algebra under the Poisson brackets
\be
\{ G_I (x) , G_J (y) \}_{PB} = {f_{IJ}}^K \d (x-y) G_K (y) \quad,
\ee
and $Q = \int_\S d^2 x \e^I (x) G_I (x)$ generate the infinitesimal 
(identity connected) gauge transformations
\be
\d A (x) = \{ Q , A (x) \}_{PB} \quad,\quad \d E (x) = \{ Q , E (x) \}_{PB} 
\quad,\quad
\d A_0^I = \dot\e^I + {f_{JK}}^I A_0^J \e^K \quad.
\ee
Hence the moduli space is the constrained phase space modulo the gauge 
transformations. Because of the second class constraint, the Gauss constraint
becomes
\be
\e^{ij}F_{ij,I} =0 \quad,
\ee
where $F = dA + A\wedge A$. Therefore the moduli space $\cal M$ is that of 
flat $G_s$ connections on $\S$.
 
The PB algebra of the second-class constraints does not close on the 
constrained phase space, and hence
the PB (\ref{pob}) will not induce a good PB on $\cal M$.
This can be remedied by introducing a second PB, the Dirac bracket
\be
\{ F , G \}^* = \{ F , G \}_{PB} - \{ F , \th_\a \}_{PB} \D^{\a\b}
\{\th_\b , G \}_{PB} \quad,
\ee
where
\be
\D^{\a\b}\{\th_\b , \th_\g \}_{PB} = \d^\a_\g \quad,
\ee
and $\a,\b,\g$ denote discrete and continious indices. 
The DB is well defined on the constraint surface, since 
$\{ \th_1 , \th_2 \}^* =0$.
The basic Dirac brackets are
\be
\{ A_i^I (x) , A_j^J (y) \}^* = \e_{ij}\eta^{IJ}\d (x-y) \label{bdb}
\ee
which will induce a symplectic structure on $\cal M$. 
The DB (\ref{bdb}) induces a symplectic 2-form
\be
\Om = \int_\S str ( \d_1 A \wedge \d_2 A ) \label{absf}
\ee
which is the supergroup generalization of the Atiyah-Bott symplectic form 
\cite{ab}.  

\section{Holonomies and the moduli space}

The standard approach for determining the moduli of flat connections is to use
the holonomies. We will use the same approach for the case of super Lie groups,
so let $U$ be a homomorphism from the fundamental group of $\S$, $\pi_1 (\S)$,
to $G_s$ defined by 
\be
U(\g) =U(s=2\p) \quad,\quad {dU(s)\over ds} = A_\m (s){dx^\m \over ds} U(s) 
\quad,\label{hol2}
\ee
where $x^\m (s) :[0,2\p] \to M$ parametrizes the loop $\g$, 
and $U(0)=Id$. Since we use a matrix 
representation
for $g_s$, which are $(m+n)\times (m+n)$ matrices, the corresponding $G_s$
will be given by $(m+n)\times (m+n)$ super matrices
\be
g = \left(\begin{array}{cc}A & B \\ 
     C & D \end{array} \right) \quad,
\ee
where $A,D$ are $m\times m$ and $n\times n$ matrices respectively, with 
entries in $B_N^+$, 
while $B$ and $C$ are $m\times n$ and $n\times m$ matrices respectively,
with entries in $B_N^-$. Given a
$B(g_s)$ one can always obtain a super Lie group by the exponential map 
$g=e^X$, but as in the case of ordinary Lie groups, one can have $G_s$ with
elements $g \ne e^X$.

The $\p_1 (\S)$ for a Riemann surface $\S$ is generated by the 
canonical cycles of $\S$,  $(a_i , b_i)$, $i=1,...,g$, 
where $g$ 
is the genus of $\S$. The generators satisfy a constraint
\be
\prod_{i=1}^{g} a^{-1}_i b_i^{-1} a_i b_i = 1 \quad.
\ee
In the case of the torus the constraint becomes $a_1 b_1 = b_1 a_1$, so that 
the 
corresponding holonomies satisfy
\be
U_1 U_2 = U_2 U_1 \quad.
\ee
Since a holonomy transforms by conjugation under  gauge 
transformations, the moduli space will be determined by commuting pairs 
$(U_1 , U_2)$ from $G_s$ modulo conjugation, i.e.
\be
(U_1 , U_2 ) \sim (\cs U_1 \cs^{-1}, \cs U_2 \cs^{-1}) \quad, \quad 
\cs \in G_s \quad.
\ee

\section{$OSp(m|2n)$ supergroup}

We now concentrate on the $OSp(m|2n)$ Lie group case. It can be defined as a 
group of super matrices $M$ 
\be
M^{st}H M = H = diag (I_m , C_{2n}) \label{ospm} 
\ee
where $C^T = -C$ and $C^2 = -I_{2n}$. If we label $M$ as
\be
M = \left( \begin{array}{cc} a &  \xi \\
          \chi &  A \end{array}\right)
\ee
and use 
\be
M^{st} = \left( \begin{array}{cc} a^T &  \chi^T \\
          -\xi^T &  A^T \end{array}\right) \quad,
\ee
then the relations (\ref{ospm}) imply
\bea
a^T a + \chi^T C \chi &=& I_m \label{om} \\
a^T \xi + \chi^T C A  &=& 0   \label{fer}\\
-\xi^T \xi + A^T C A    &=& C \quad.\label{spn}
\eea
These relations can be solved by expanding the matrices $a,A,\chi,\xi$ in
the Grassman algebra $B_N$ basis as
\be
X = X_0 + X_1 \th^1 + X_2 \th^2 + ... + X_N \th^N \quad,
\ee 
where $\th^k = \th_{i_1} \cdots \th_{i_k}$ and we have supressed the $i_l$
indices on $X_k$. Note that for physics applications $N \ge 2mn$, 
otherwise $N$ is an arbitrary natural number.

The relations (\ref{om}) and (\ref{spn}) imply 
\be
a_0^T a_0 = I_m \quad,\quad A_0^T C A_0 = C \quad, 
\ee
so that $a_0 \in O(m)$, while $A_0 \in Sp(2n)$. This implies that $a$ and $A$
are always invertible, so that the relation (\ref{fer}) gives
\be
\xi = -(a^T)^{-1}\chi^T C A \quad,\label{ksi}
\ee
which means that
there is only one independent fermionic matrix. Hence if $\chi =0$, then
$\xi =0$ and therefore $a \in O(m)$ while $A \in Sp(2n)$, so that one obtains 
the maximal bosonic subgroup $O(m)\times Sp(2n)$.

If $M$ is a holonomy matrix $U$, we would like to know how the fermion $\chi$
transforms under the gauge transformation
\be
\tilde U = \cs U \cs^{-1} = \left( \begin{array}{cc}\tilde a & \tilde\xi \\
          \tilde\chi &  \tilde A \end{array}\right) \quad, \label{tiu}
\ee
where $\cs \in OSp(m|2n)$. If 
\be
\cs = \left( \begin{array}{cc} s &  \d \\
          \s &  S \end{array}\right)
\ee
then
\be
\cs^{-1}=\left( \begin{array}{cc}{\bar s}^{-1} &  -s^{-1}\d {\bar S}^{-1} \\
          -S^{-1}\s {\bar s}^{-1} &  {\bar S}^{-1} \end{array}\right) \quad,
\label{ins}
\ee
where $\bar s = s -\d S^{-1}\s$ and $\bar S = S -\s s^{-1}\d $.
By using (\ref{ins}), the relation (\ref{tiu}) gives
\be
\tilde\chi = \left( \s a + S\chi - \s\xi S^{-1}\s - SAS^{-1}\s \right)
{\bar s}^{-1} \quad. \label{tic}
\ee
Therefore if we want $\tilde\c =0$, then (\ref{tic}) implies
\be
 \s a + S\chi - \s\xi S^{-1}\s - SAS^{-1}\s = 0  
\quad. \label{zfc}
\ee
This can be considered as an equation for $\s$, which can be written in
components of a $B_N$ basis as
\be
\s_{2k+1} a_0 - S_0 A_0 S_0^{-1} \s_{2k+1} = -S_0 \c_{2k+1} + 
\zeta_{2k +1}(\s_1,...,\s_{2k-1})
\quad,\label{rrs}
\ee
where $\zeta_1 = 0$. Therefore (\ref{rrs}) can be solved iteratively for 
$\s$ as a
function of $a, A, \c$ and $s, S$ provided that the linear operator $\hat A$
determined by
\be
\hat A \s =  \s a_0 - S_0 A_0 S_0^{-1} \s
\ee
is invertible. It can be written as a $2mn \times 2mn$ matrix
\be
\hat A_0 = a_0^T \otimes I_{2n} - I_m \otimes S_0 A_0 S_0^{-1} \quad,
\ee 
so that invertability is equivalent to $\det \hat A_0 \ne 0$. 
Note that
\be
{\hat S}^{-1} \hat A_0 {\hat S} = a_0^T \otimes I_{2n} - I_m \otimes A_0 
\ee
where $\hat S = I_m \otimes S_0 $, so that
\be
\det ( a_0^T \otimes I_{2n} - I_m \otimes S_0 A_0 S_0^{-1}) =
\det ( a_0^T \otimes I_{2n} - I_m \otimes  A_0 ) \quad. \label{fec}
\ee

In the case when $\det\hat A_0 \ne 0$ one
can determine all the components of $\s$ and hence set all of the components of
$\tilde\chi$ to zero. This then means that the holonomy matrix $U$ is 
equivalent to a holonomy matrix of the $O(m)\times Sp(2n)$ subgroup. If $U_1$
is such a matrix, i.e. $U_1 = diag(a,A)$ then $U_2$ must be also, since the 
commutativity implies
\be
a b = ba \quad,\quad A B = B A\quad, \quad \m a  = A\m  \quad,\quad
a\n  = \n A \quad,
\ee
where
\be
U_2 = \left( \begin{array}{cc}b &  \n \\
          \m &  B \end{array}\right)\quad.
\ee
Since $\det ( a^T \otimes I_{2n} - I_m \otimes  A ) \ne 0$ then
$\m a - A \m  =0 $ implies $\m =0$, and hence $U_2 = diag(b,B)$.
Hence
the moduli space is that of the $O(m)\times Sp(2n)$ Lie group, and there are no
fermionic moduli. 

Note that this bosonic moduli space will be larger then the 
moduli space of non-super CS based on the $O(m)\times Sp(2n)$ Lie group because
the holonomy matrices in the former case belong to the $B_N^+$ Grassman algebra
while in the latter case the holonmy matrices are real numbers. We can avoid 
these extra bosonic degrees of freedom by restricting the even part of the 
CS connection $A$ from $B_N^+$ to real numbers. 

The fermionic moduli will arise in the case when $\det\hat A_0 = 0$ and 
$\det\hat B_0 = 0$ where
\be
U_1 = \left( \begin{array}{cc}a &  \xi \\
          \c &  A \end{array}\right)
\quad,\quad
U_2 = \left( \begin{array}{cc}b &  \n \\
          \m &  B \end{array}\right)\quad.
\ee
The zero commutator implies
\be
a b + \xi\m = ba + \n\c \label{smc}
\ee
\be
 A B + \c\n = B A + \m\xi \label{lac}
\ee
\be
\c b + A \m = \m a + B\c \label{fec}
\ee
\be
a\n + \xi B = B\xi + \n A \quad. 
\ee
Let $rank (\hat A_0 )= r < 2mn$, so that $r$ components of
$\c$ can be set to zero. Although the non-zero components are not gauge 
invariant, they cannot be set to zero by a gauge transformation. We will
show in the following sections  how to define the gauge invariant fermion 
components, but for our present purposes it sufficient to know that the
number of non-zero fermion components is gauge invariant. 
If $rank (\hat B_0 )= r^\prime $, then one can prove that
$r=r^\prime$. From (\ref{smc}) and (\ref{lac}) it follows 
\be
[a_0 , b_0 ] = 0 \quad,\quad [A_0 ,B_0 ]=0 \quad,\label{bca}
\ee
while from (\ref{fec}) it follows
\be
 A_0 \m_1 - \m_1 a_0 =  B_0\c_1 -\c_1 b_0 \quad.\label{feca}
\ee
The relation (\ref{feca}) can be rewritten as
\be
\hat A_0 \hat\m_1 = \hat B_0 \hat\c_1 \label{fecb}
\ee
From (\ref{bca}) it follows that $[\hat A_0 , \hat B_0 ] =0$ and hence 
these matrices can be simultaneosly diagonalised (or put in the  
block-diagonal form). If $r^\prime > r$, then from \ref{fecb} it follows
that $r^\prime$ components of $\c$ vanish, which is in contradiction with the
assumption that only $r$ components of $\c$ vanish. Hence $r^\prime \le r$.
If $r^\prime < r$ then by reversing $\m$ and $\c$ and by the same argument 
it follows $ r \le r^\prime$ so that $r = r^\prime $. 
Hence the number of the independent fermionic (odd) moduli           
is equal to $2(2nm - r)$.

The total number of real (complex) parameters for the fermionic phase-space
is $(2nm - r )e^{2nm}$. Physics expectation is $(2nm -r)4nm$, i.e. only
$\c_1$ and $\m_1$. This happens when $\psi \in B_1$.
 
\section{Symplectic structure of the moduli space}

We now discuss some aspects of the symplectic structure on the moduli space
which will be useful for an understanding of the results we have derived so 
far.
A natural symplectic structure on $\cal M$ can be induced by the symplectic 
structure (\ref{absf}) on the space of connections. By using the definition of 
$U(\g)$ and (\ref{bdb}) one obtains 
\bea
\{ U_A^B (\g) , U_C^D (\s) \}^* &=& 2s(\g,\s)\sum_{E,F,G,H}
(-1)^{[(|B|- |E|)(|A| - |E|) + (|C|-|H|)(|D| -|H|)]}\nonumber\\
& & U_A^E (\g_i) (T^I)_E^F 
U_F^B (\g_f) U_C^G (\s_i)(T_I)_G^H U_H^D (\s_f) 
\quad, \label{hdb}
\eea
for loops with at most one intersection, where $s(\g,\s)$ is the intersection 
number \cite{urr}. From this algebra one can find in principle the PB algebra
of the coordinates on the moduli space and hence the symplectic form. However,
this is beyond the scope of this paper, and instead we will analyse a 
simpler algebra, induced by the homogenous connections, since it will give
a very good idea of the moduli space, especially of the fermionic
moduli. 

Consider the holonomies which are the exponentials of the Lie super algebra, 
i.e. the subgroup $g_s = \exp X$. Then $U_k$ are   
given by 
\be
U_k = \exp (2\pi A_k^a J_a + 2\pi\psi_k^\a Q_\a ) \quad,
\ee
where $A_k$ and $\psi_k$ are constants on $\S$. These correspond to the
homogenous sector of the moduli space, for which the corresponding connections
can be written as
\be
A = A_1 d\th + A_2 d\phi =  (A_1^a J_a + \psi_1^\a Q_\a)d\th +
(A_2^a J_a + \psi_2^\a Q_\a)d\phi \quad,
\ee
where $\th,\phi \in [0,2\pi ]$ parametrize the $a,b$ cycles of the torus.
Hence $[U_1 , U_2]=0$ is equivalent to $[A_1 , A_2 ]=0$ which gives the
constraints
\bea
G^a = {f_{bc}}^a A_1^b A_1^c + {f_{\a\b}}^a \psi_1^\a \psi_2^\b =0 \quad,
\label{hbc} \\
G^\a = {f_{a\b}}^\a ( A_1^a \psi_2^\b -  A_2^a \psi_1^\b )=0 \quad.\label{hfc}
\eea
These are first-class constraints and form an $osp(m|2n)$ algebra under the
Poisson brackets (\ref{bdb}), which for the homogenuous sector become
\be
\{ A_k^a , A_j^b \}^* = \e_{kj}\eta^{a b} \quad,\quad
\{ \psi_k^\a , \psi_j^\b \}^*  = \e_{kj}C^{\a \b} 
\quad.\label{hfdb}
\ee
The relations (\ref{hfdb}) imply that the space $A,\psi$ is the usual 
phase space,
i.e. half of the parameters are coordinates and the other half are the momenta.

According to the theory of solving the constraints in phase
space \cite{gf,bf,he}, the constraints (\ref{hbc}) and (\ref{hfc}) can be 
solved in the following 
manner. In the bosonic sector, if we require that $A_k$ are
ordinary numbers, then the bosonic constraint implies that $A_k^a J_a$
belong to the same abelian subgroup, and hence $A_k^a = {\cal A}_k c^a$.
In the fermionic sector, let
$r$ be the number of independent fermionic constraints $G^{\tilde\a}$. 
$r$ will be given by the rank of the matrix $ c^a {f_{a\a}}^\b $,
and hence one can
impose $r$ fermionic gauge-fixing conditions $\c^{\tilde\a} = 0$ such that
\be
\det ||\{ G^{\tilde\a},\c^{\tilde\b}\}^* || \ne 0 \quad.\label{ffpd}
\ee
This gives
$4nm -r -r = 2(2nm -r)$ independent fermionic phase-space coordinates, while
in the bosonic sector there at most two \footnote{There may be a further
constraint for some Abelian subgroups} independent phase space coordinates
${\cal A}_1$ and ${\cal A}_2$. 

Therefore the condition for the existence of the fermionic moduli
for the exponential sector becomes
\be
\det || c^a {f_{a\a}}^\b || = 0 \quad.\label{efm} 
\ee
Note that our analysis for the exponential sector
is general, and applies to any super Lie group. Hence one finds all 
inequivalent
(with respect to conjugation)  Abelian subgroups of the maximal bosonic 
subgroup, and checks the condition (\ref{efm}). There is also an additional
constraint 
\be
{f_{\a\b}}^a \psi_1^\a \psi_2^\b =0 \label{afc}
\ee
coming from the bosonic constraint (\ref{hbc}). The constraint (\ref{afc}) 
will serve to determine
the gauge-fixing functions $\c^{\tilde\a} =0$, i.e.
$\c^{\tilde\a}$ will be linear in $\psi_k$ such that (\ref{afc}) is satisfied
as well as the condition (\ref{ffpd}) (see section 7).

Also note that the matrix
$J_\a^\b = c^a {f_{a\a}}^\b $ is the representation of the generator of the
Abelian subgroup to which the bosonic holonomy belongs, and it is the 
fermion-fermion block in the super-adjoint representation of the super Lie
algebra $(T_I )_J^K = {f_{IJ}}^K$. In the $osp(m|2n)$ case 
the condition $\det\hat A_0 =0$ applies to any holonomy, and in the 
exponential sector it is equivalent to (\ref{efm}). 

\section{$OSp(1|2)$ case}

We now present explicit constructions in the simplest
non-trivial case of the $OSp(1|2)$ supergroup.
In this case we have $m=n=1$, so that $a$ is a Grassman number
and $A$ is an $Sp(2)= SL(2)$ matrix, and we take a $B_2$ Grassman algebra. 
The $osp(1|2)$ algebra can be represented by matrices satisfying
\bea
[ J_a ,J_b ] &=& {\e_{ab}}^c J_c \quad,\\
{[} J_a ,Q_\a ] &=& {(\s_a)_\a}^\b Q_\b  \quad,\\ 
 \{ Q_\a ,Q_\b \} &=& (\s^a)_{\a\b} J_a \quad,\label{osp}
\eea
where $a = 0,1,2$ and the matrices ${(\s_a)_\a}^\b$ are 
\be
\s_0 = \left( \begin{array}{cc}0 &  1 \\
          -1 &  0 \end{array}\right)\quad,\quad
 \s_1 = \left( \begin{array}{cc} 1 &  0 \\
          0 &  -1 \end{array}\right)\quad,\quad
\s_2 = \left( \begin{array}{cc} 0 &  1 \\
          1 &  0 \end{array}\right) \quad.
\ee
The indices $a$ and $\a$ are raised and lowered by $\eta_{ab}=diag(-1,1,1)$
and $C_{\a\b}= \e_{\a\b}$. The relevant matrix representation of the 
generators of the $osp(1|2)$ algebra is given by
\be
J_a = \left( \begin{array}{cc}1 &  0 \\
          0 &  \s_a \end{array}\right) 
\quad,\quad
Q_\a = \left( \begin{array}{cc}0 &  -c_\a^T C \\
          c_\a &  0 \end{array}\right)
\ee
where $c_1^T =(1,0)$ and $c_2^T = (0,1)$. The exponential map gives only a
subgroup of the full group. 

For the group have $a_0=\pm 1$, and inequivalent bosonic Abelian subgroups 
are given by
\be
\pmatrix{1 & 0  \cr 0 & e^{\phi \s_0}}\quad,\quad 
\pmatrix{1 & 0  \cr 0 & e^{v \s_1}}\quad,\quad 
\pmatrix{1& 0  \cr 0 & e^{c (\s_0 + \s_2)}} 
\ee 
where $\s_0$, $s_1$ and $\s_+ = \s_0 + \s_2$ generate non-equvalent abelian
subalgebras of $sl(2)$.

Since any element of $SL(2)$ can be written as $\pm e^X$, the bosonic conjugacy
classes  will be given by pairs
\be
(U_1 , U_2 ) = \left( \pmatrix{a_0 & 0 \cr 0 & \e_1 e^X} , 
\pmatrix{b_0 & 0 \cr 0 & \e_2 e^Y} \right) \quad,
\ee
where $\e_k = \pm 1$ and $X$ and $Y$ belong to the same Abelian subalgebra of
$sl(2)$ which is not the $so(2)$ subalgebra. When $X,Y \in so(2)$, 
then
\be
(U_1 , U_2 ) = \left( \pmatrix{a_0 & 0 \cr 0 & e^X} , 
\pmatrix{b_0 & 0 \cr 0 & e^Y} \right) \quad.
\ee
Hence there will be $2\cdot 16 + 4 = 36$ sectors of bosonic conjugacy classes.

The fermionic moduli will exist for holonomies which satisfy the
condition $\det\hat A_0 =0$, which becomes
\be
\det ( a_0 I_2 - A_0 ) = 0 \quad.
\ee
This is solved by
\be
a_0 = \pm 1 \quad,\quad A_0 =   \left( \begin{array}{cc}\pm 1 &  c \\
          0 &  \pm 1 \end{array}\right)\quad.
\ee
Note that $A_0 = \pm e^L$, where $L= c \s_+$, i.e. $e^L$ belongs to 
the $GL(1)$ subgroup of $SL(2)$.
Hence $rank \hat A_0 = 1$, and therefore there will be $4 -1-1 =2$ fermionic
moduli. The corresponding holonomy matrices are given by
\be
U_k =  \left( \begin{array}{cc} \pm 1 &  \xi_k \\
                                 \c_k & \pm e^{L_k} \end{array}\right) 
\label{sgf}
\ee
where $\c_k = (0 , \m_k )^T$ and $\xi_k = -a_0 \c_k^T C A_0$.
Hence there are $2\cdot 2 = 4 $ different sectors of conjugacy classes 
$(U_1,U_2)$ which contain fermions.
One can have more general solutions by replacing $A_0$ by 
$A_0 + A_2 \th_1\th_2$ where $ A_0^T C A_2 + A_2^T C A_0 =0$, but as we 
discussed, we only consider physical connections.
Since $U_k$ belong to the parabolic conjugacy class of $SL(2)$, we will
have $c_1^2 + c_2^2 = 1$ \cite{wmn}.

For the sector where $U = \exp X$, we have 
\be
U_k = \exp 2\pi ( {\cal A}_k \s_+ + \psi_k^\a Q_\a )
\ee
where $\s_+ = \s_0 + \s_2$ and
\be
\{ {\cal A}_1 , {\cal A}_2 \}^* = 1 \quad,\quad \{\psi_1^\a , \psi_2^\b \}^* = C^{\a\b}
\ee
The fermionic constraint yields
\be
{\cal A}_1 \psi_2^+ - {\cal A}_2 \psi_1^+ = 0 \quad,\label{fc}
\ee
where $\psi^\a = (\psi^+ \, \psi^- )$. By inserting the fermionic constraint
into the bosonic constraint, one obtains
\be
{\cal A}_1 \psi_2^- - {\cal A}_2 \psi_1^- = 0 \quad.\label{gfc}
\ee
This condition represents a consistent gauge-fixing choice, so that
there exists 
a canonical transformation $\psi_k \to \tilde\psi_k$ such that 
\be
\tilde\psi_1 = (\Pi \,, \tilde\Pi )^T \quad,\quad
\tilde\psi_2 = (\tilde\psi\,, -\psi )^T \quad, 
\ee
and the constraints (\ref{fc}) and (\ref{gfc}) become equivalent to 
$\tilde\Pi =0$ and $\tilde\psi =0$ \cite{he}. With such a choice of
fermionic variables, and the fact that ${\cal A}_1 = p$ and ${\cal A}_2 = q$ 
where
$\{p , q \}^* = 1$, one can now express $c_k$ and $\c_k$ in terms of basic
canonical variables, and work out their PB. Constructing the canonical 
transformation $\psi_k\to \tilde\psi_k$ may not be easy, but one can show
that
\be
U_1 = \exp 2\pi \left( p \s_+ + {p\over q}c^\a (p,q) Q_\a \psi \right) 
\quad,\quad
U_2 = \exp 2\pi \left( q \s_+ + c^\a (p,q) Q_\a \psi \right) \quad, \label{crh}
\ee
where $c^\a$ can be determined from the requiriment that 
$\psi_2^\a = c^\a (p,q) \psi + \cdots$ where $\cdots$ represent terms
which vanish on the constraint surface.

For the non-exponential sector one can use the holonomy matrix algebra, but
in some cases it is possible to obtain explicit expressions in terms of the
basic canonical variables. For example, in the sector $a_0 =1$, 
$A_0 = -e^X$ (which has no fermionic moduli)
one can construct the matrices $U(\phi)$ and $U(\th)$ as
\be
U(\phi) = diag(1,R(\phi/2))\exp ({\cal A}_1 \s \phi + \psi_1^\a Q_\a \phi) 
\quad,
\ee
where $R(\phi/2)$ is the rotation matrix for angle $\phi/2$, and $U(\th)$
is given by replacing $\phi$ with $\th$ in $U(\phi)$ and ${\cal A}_1$ with
${\cal A}_2$,
so that $U(0)=Id$ and $U(2\pi)= U_{1,2}$. Hence the corresponding
flat connections are given by
\be
A_1 = U^{-1}(\phi)\pa_\phi U(\phi)\quad,\quad 
A_2 = U^{-1}(\th)\pa_\th U(\th)
\ee
and they are not constant on $T^2$, but they are expressed in terms of 
constant moduli with simple PB. The moduli ${\cal A}_k$ satisfy the same 
constraints as in the homogeneous case, and the same results apply for them.
However, this trick does not work for the holonomies with $a_0 = -1$ and
$A_0 = - e^X$.

Since $c_1^2 + c_2^2 = 1$, the symplectic structure for the bosonic moduli
becomes degenerate. In the exponential sector this is equivalent to 
$p^2 + q^2 = 1$ so that
$p = \cos t$, $q= \sin t$ which suggests a dynamical interpretation
as a harmonic oscillator with energy $E=1/2$.

The degeneracy of the bosonic symplectic structure for holonomies with 
fermions appear because the group is small. For bigger groups, there is
generically
no such degeneracy. For example, for $OSp(2|2)$ case we have $a_0$ is an $O(2)$
matrix while $A_0$ is an $SL(2)$ matrix. In the sector where
$a_0 \in SO(2)$, so that $a_0 = R(\phi)$, the condition for the existence 
of the fermionic moduli is
\be
\det\hat A_0 = \det (a_0^T \otimes I_2 - I_2 \otimes A_0 )= 
(2\cos\phi - tr\, A_0)^2 = 0 \quad,
\ee
Hence $A_0$ belongs to the $SO(2)$ subgroup of the $SL(2)$, and therefore
the bosonic conjugacy classes
which have the fermionic moduli are
\be
U_1 = diag(R(\phi_1), R(\pm\phi_1)), U_2 =diag(R(\phi_2), R(\pm\phi_2))
\quad.
\ee
The corresponding symplectic form is $d\phi_1 \wedge d\phi_2$, and it
is nondegenerate. Since $rank\hat A_0 = 2$, there will be $4$ fermionic 
moduli. Also in this sector one can use the
exponential map to construct the moduli with canonical PB.

\section{Conclusions}

The structure of the total moduli space is essentially determined by
the bosonic moduli associated with the maximal bosonic subgroup. 
This is reminiscent 
of the fact that  the supergroups have no extra topology associated to
the fermionic sector \cite{rp}. Note that the bosonic moduli
space can have quite a complicated topology, (e.g.
non-Hausdorff, see \cite{ml}). It would be 
interesting to see how the bosonic topology affects the
fermionic moduli topology.

The fermionic moduli appear only in special sectors of the bosonic moduli 
space. In the $OSp(m|2n)$ case this condition is given by the vanishing
of a determinant coming from the c-number part of the holonomies associated
to Abelian subgroups of the maximal bosonic subgroup. In the
exponential sector this condition becomes a vanishing determinant 
of the Abelian subalgebra generator in the fermion-fermion block of the
super-adjoint representations. This is true in the general case for the 
exponential sector. One can use the homogeneous connections to describe the
exponential sector, and the example given in the section 7 for a 
non-exponential
holonomy shows how the constant connections could be used to describe
the non-exponential sector by the moduli which have canonical PB.

The fermionic canonical moduli $\psi$ are not gauge invariant, 
and one can obtain gauge invariant
fermionic moduli $\tilde\psi$ via canonical transformation. The non-exponential
sector can be examined through the algebra of holonomies, and one can use
the results for the $OSp$ case to study the algebra. It is not clear how
the supertraces in \cite{urr} capture the properties of the
fermionic moduli, since the relevant holonomies have supertraces 
which are
c-numbers, and not elements of $B_N^+$. To understand this point it would be 
helpful to study the algebra of the holonomy matrices (\ref{sgf}).  

Another interesting question is how to represent $U$ in 
terms of basic canonical variables. This can be worked out in the
exponential sector, and in order to
generalise these results to the non-exponential
sector and higher genus surfaces one would need a procedure for 
reconstructing the
connection from the holonomies in the case of non-compact groups and
higher-genus surfaces. Note that such a construction exists for compact and
semisimple groups and surfaces of arbitrary genus \cite{ks}.
Also in principle the explicit correspondence between holonomies and general
connections described in \cite{mp} could be adapted to the situation at hand. 

\section*{Acknowledgements} 
A. M. was supported by the grant 
PRAXIS/BCC/18981/98 from the Portugese Foundation for Science and Technology
(FCT). R. P. was supported by Financiamento Plurianual provided by the FCT.

\end{document}